\documentclass[12pt]{iopart}
\usepackage[T1]{fontenc}
\usepackage[utf8]{inputenc}

\normalsize

\usepackage{capt-of}
\usepackage{caption}
\usepackage{tikz}
\usepackage{pgfplots}

\expandafter\let\csname equation*\endcsname\relax
\expandafter\let\csname endequation*\endcsname\relax
\usepackage{amsmath}
\usepackage{graphicx}
\usepackage{bm}
\usepackage{longtable}
\usepackage{amssymb}
\usepackage{tensor}
\usepackage[colorlinks=true]{hyperref}
\usepackage{sidecap}
\usepackage{wrapfig}
\usepackage{soul}
\usepackage{orcidlink}
\usepackage{mathenv}

\usepackage{aas_macros}

\pgfplotsset{compat=1.18}
\begin{document}
\title[Detecting classical nova-like explosions]{Detecting classical nova-like explosions with LISA}
\author{Ethan McKeever$^1$\orcidlink{0009-0002-8956-7849}, Shu Yan Lau$^1$\orcidlink{0000-0002-8239-0174}, Hang Yu$^1$\orcidlink{0000-0002-6011-6190}, and Neil J. Cornish$^1$\orcidlink{0000-0002-7435-0869}}
\address{$^1$eXtreme Gravity Institute, Department of Physics, Montana State University, Bozeman, Montana 59717, USA}
\ead{ethanmckeever@montana.edu}
\ead{hang.yu2@montana.edu}

\begin{abstract}

Gravitational waves from close binary white dwarfs will form the bulk of measurements obtained by the Laser Interferometer Space Antenna (LISA). Previous studies have highlighted the importance of including the effects of steady-state mass transfer in waveform models as many individually resolvable white dwarf binaries will be interacting during the LISA observation window. However, few studies have considered the effect of novae on gravitational wave observations and parameter estimations. We fill in this gap by analyzing the detectability of novae in these systems and the biases in physical parameters when these bursts are not considered. We model a nova burst as a rapid loss of mass from the accretor that suddenly shifts the gravitational wave frequency. We analytically predict the signal-to-noise ratios for direct detection of the nova and where bias from mismodeling becomes greater than the statistical uncertainty. We also consider comparison of two halves of future LISA data to detect bursts in a model agnostic way. Our work has implications for identifying and classifying individual nova events as well as constraining the galactic nova rate throughout the entire galaxy, a challenging task for optical surveys.

\end{abstract}

\maketitle

\section{Introduction}

Population studies on double white dwarf binaries (DWDs) have estimated that more than 60 million binaries exist in the Milky Way, emitting gravitational waves (GWs) at frequencies that will be probed by
the Laser Interferometer Space Antenna (LISA, \cite{Lamberts:19}). The vast majority of these systems exist at the low end of the frequency range (only $\sim 6 \%$ of galactic DWD binaries are expected to have frequencies between 0.1 and 10 mHz) and will serve as a source of background confusion noise \cite{Bender:97, Ruiter:10, Karnesis:21}. However, an estimated 10,000 will be individually resolvable for a four-year mission timeline. These resolvable systems predominantly exist near the higher end of the galactic DWD frequency distribution with a median between $1-2$ mHz \cite{Lamberts:19} and an estimated 2,720 will have negative chirps greater than 0.1 frequency bins per year due to mass transfer effects \cite{Kremer:17}. Many of these systems are expected to also be observable with electromagnetic methods \cite{Nelemans:04}. It is likely that some of these systems will experience nova bursts during the LISA observation period due to the buildup of semi-degenerate matter on a degenerate core \cite{Giannone:67, Rose:68, Starrfield:71a, Starrfield:71b, Kaplan_2012}.  For systems with rapid mass transfer and high mass primary stars the recurrence time between bursts may be as short as a few months \cite{Yaron:05}. These bursts will act as rapid shifts to the GW frequency and should be included in LISA data analysis to prevent binary parameter estimation bias (especially stealth biases) and to accurately perform a global fit \cite{Cornish:05, Vallisneri:09, Littenberg:20, Littenberg:23}.

 There is great uncertainty as to how these bursts will affect DWD orbits as they have not yet been observed in DWDs using electromagnetic observations and accurate period shifts for classical and recurrent novae have proven challenging to obtain \cite{Schaefer:21, Schaefer:23}. Isotropic mass loss with conserved angular momentum will lead to orbital expansion while dynamical friction within the expanding nova shell may cause orbital shrinkage leading to a positive overall frequency shift \cite{Shen_2013, Shen:15}. If binaries are able to survive these nova bursts, they may exist for significant periods as stably mass transferring systems \cite{Marsh:04}. Within these systems, a donor helium WD can retain a thin shell of hydrogen of $\sim 10^{-3}M_{\odot}$ \cite{Fuller:13} and power up to 1,000 classical nova-like explosions \cite{Yaron:05}. After the donor is depleted of hydrogen-rich material, similar thermonuclear explosions can still occur as helium shell flashes \cite{Bildsten:07, Shen_2009}. Novae/helium flashes can significantly perturb the orbital frequency of a binary. For classical novae, the observed fractional change in period has been as large as $10^{-3}$ \cite{Schaefer:20_b, Schaefer:20}.

Future GW observations will also shed light on proposed but unobserved nova classifications. We will be able to observe potential progenitors of type ``.Ia'' supernovae, helium flashes one-tenth the luminosity of standard type Ia supernovae. 
While the last flash triggering the .Ia is likely too rare to be directly observed, \cite{Bildsten:07} predicts multiple weaker but much more frequent precursor flashes. A few of these flashes may occur within the galaxy during LISA's observational period. Another exciting measurement would be observations of tidal novae: runaway nuclear reactions caused by the deposition of tidal heating in the outer layers of carbon-oxygen white dwarfs \cite{Fuller_2012}. These novae would clearly stand out from other bursts as they are expected to precede the onset of mass transfer.

Orbital evolution equations from \cite{Marsh:04} require mass transfer rates of $10^{-7} - 10^{-8}$ solar masses per year, depending on tidal efficiency and mass ratio, for anti chirping of 0.1 frequency bins per year. For an accretor of 0.65 solar masses and a core temperature of $3\times10^7$ K, these transfer rates lead to burst recurrence periods of 254 - 10,200 years \cite{Yaron:05}. For the 2,720 binaries predicted by \cite{Kremer:17}, this leads to an expectation of 1 - 43 galactic novae from LISA binaries for a four year observation time. 

When performing parameter estimation in this work we make use of a noise-free yet noise weighted likelihood, which allows for rapid results via a heterodyned likelihood \cite{Cornish:13}. This method returns realistic spreads in parameter estimations with the center of the parameter distribution lining up exactly with the injected values when recovering values using the injected model. In section \ref{sec:Waveform} we present our waveform model and Fisher information matrix. In section \ref{sec:Detection Thresholds} we present thresholds for detections of nova bursts. In section \ref{sec:Stealth Bias} we discuss mismodeling bias in parameter estimation when the effects of the burst are not included. In section \ref{sec:stealth bias 2} we discuss how to identify bursts and similar phenomena in a model agnostic way for future observations. In section \ref{sec:Discussion} we discuss our results and future work. 

We make all relevant programs used in the preparation of this paper available to interested readers in the form of a \href{https://github.com/Ethan-McKeever/Detecting-classical-nova-like-explosions-with-LISA}{GitHub repository}.

\section{Burst Model and Waveform} \label{sec:Waveform}
\begin{figure}
    \centering
    \includegraphics[width=0.95\linewidth]{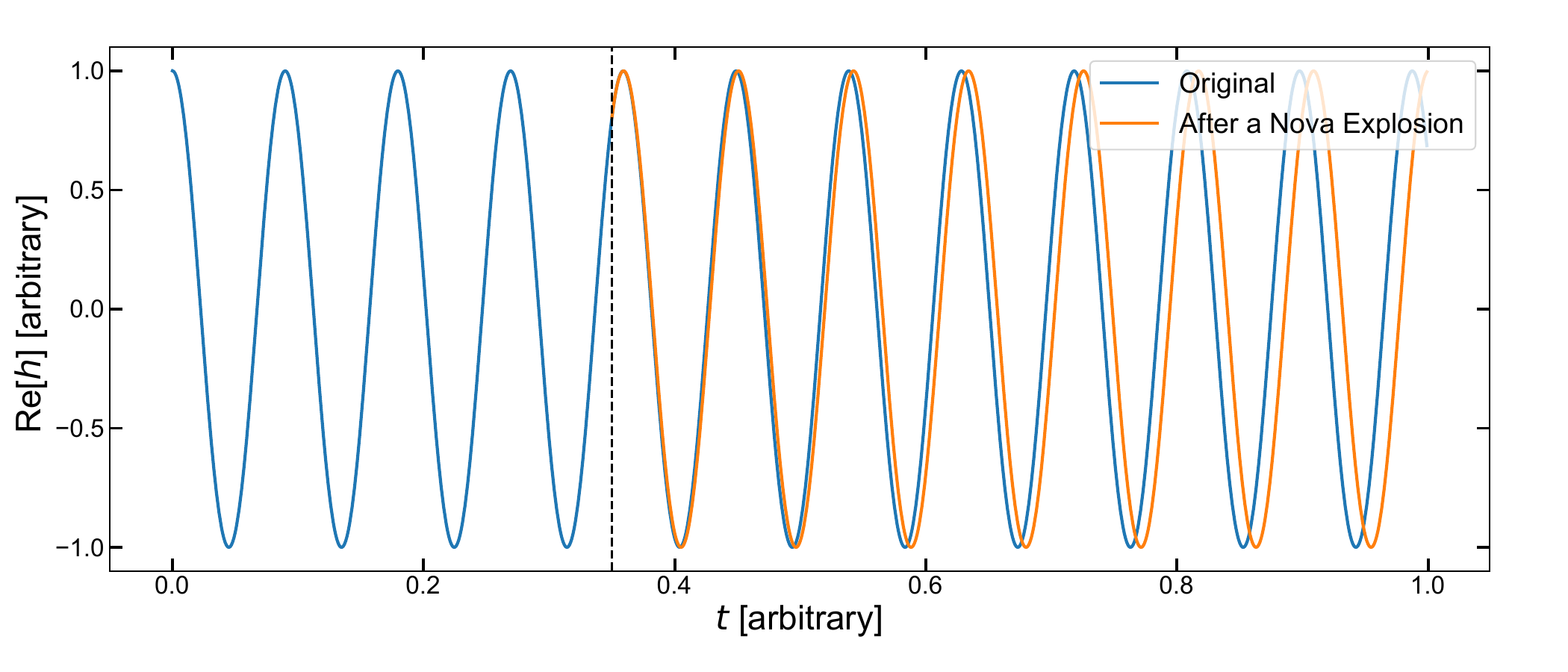}
    \caption{A schematic illustrating the effect of a nova-like explosion (indicated by the black vertical line) in an accreting DWD. The sudden ejection of up to $10^{-4} M_\odot$ of mass can cause the orbit to expand, reducing the GW frequency by a similar fractional amount.  
    (The effect is shown here with exaggerated shift of $\Delta f/f = 0.02$ for illustration purposes so that it's visible within a few cycles.)
    }
    \label{fig:h_w_nova}
\end{figure}

We model a nova burst that induces an instantaneous frequency shift $\Delta f$ at a single given time $t_b$ which is scaled by the total observation time $T_{\rm{obs}}$ and ranges between 0 and 1. These choices are motivated by a theoretical analysis of classical novae that showed that most bursts have a total time of mass loss on the order of a few days and a recurrence period longer than the planned mission lifetime for LISA \cite{Yaron:05}. We also constrain the frequency shift such that the waveform phase is continuous across the event and limit our considerations of the burst to the frequency shift; we do not consider adjustments to the strain amplitude or frequency derivatives (to be justified later). Similar models are applied to describe the resonant excitations of stellar oscillation modes in the inspiral waveform \cite{Flanagan_2007, Yu:17}. An example waveform with an exaggerated burst is shown in Fig. \ref{fig:h_w_nova}

Our analysis in this work is applicable to nova bursts that cause either an overall positive or negative frequency shift. The results are symmetric around zero and only depend on the magnitude of $\Delta f$. As an order of magnitude estimate to set our prior range we consider a conservation of angular momentum argument as follows. We start with the orbital angular momentum of a quasi-circular binary:
\begin{eqnarray}
    J_{\rm{orb}} = \left(\frac{G^2}{M\pi f} \right)^{1/3}M_1M_2,
\end{eqnarray}
where $G$ is Newton's constant, $M$ is the total mass, $M_1$ and $M_2$ are the primary and donor masses (with $M_1 > M_2$), and $f$ is the GW frequency. We then perturb the system by ejecting some mass $M_1 \rightarrow M_1-\Delta m$ with $J_{\rm{orb}}$ conserved. We compensate by including a shift in frequency: $f \rightarrow f + \Delta f$. To first order in $\Delta m/M_1$, this grants:
\begin{equation}
    \frac{\Delta f}{f} \simeq \frac{\Delta m}{M} - \frac{3\Delta m}{M_1} = -\frac{\Delta m}{M_1}\frac{2+3q}{1+q},
    \label{eq:df_f_est}
\end{equation}
where $q = M_2/M_1$ is the mass ratio. Theoretical modeling of nova explosions shows expected values of $\Delta m/M_1 \in[4 \times10^{-8},10^{-3}]$ \cite{Yaron:05, Townsley:04} while observational data of period shifts for classical novae offer similar bounds $|\Delta f|/f \in[2\times10^{-6},2\times10^{-3}]$ \cite{Schaefer:23}. Notably, five of the fourteen observed frequency shifts are positive, including the shift with the largest magnitude. These results show dynamical friction between the donor and ejecta may overpower the above angular momentum effects, an outcome studied in DWDs by \cite{Shen:15}.

When the accretor loses mass, the frequency evolution rate $\dot{f}$ should also change by a fractional amount similar to Eq. (\ref{eq:df_f_est}). However, as we will see later, the fractional uncertainty on the frequency derivative is about a few percent for a typical DWD with signal-to-noise ratio (SNR) slightly above 10. Meanwhile, the amplitude of the GW signal has minimal correlation with the other parameters. Therefore, we ignore the impact of mass loss on the frequency derivative and strain amplitude.

We consider a slowly evolving binary and Taylor expand the waveform to second order in time. We use dimensionless parameters $\alpha = T_{\rm{obs}}f$, $\beta = T_{\rm{obs}}^2\dot{f}$, and $\gamma = T_{\rm{obs}}\Delta f$. We also use a dimensionless time $t \in [0,1]$. For much of this work we consider a binary with $f = 9.7$ mHz and $\dot{f} = -3.1\times10^{-15}$ Hz/s, corresponding roughly to a $q=1/3$ binary of cold WDs undergoing mass transfer using the fit from \cite{Eggleton:83} and evolution equations presented in \cite{Marsh:04}. These grant values of $\alpha =1.22\times10^6$ and $\beta = -49.5$. 

Intrinsic information for the binary is contained in $\beta$ as \cite{Marsh:04}:
\begin{equation} \label{eq: betadef}
    \frac{\beta}{3T_{\rm obs} \alpha} = -\frac{\dot{J}_{\rm GR}}{J_{\rm orb}}- \frac{k M_1 R_1^2}{\tau_s J_{\rm orb}}\omega + \left[1-q-\sqrt{(1+q)r_{\rm h}} \right]\frac{\dot{M}_2}{M_2},
\end{equation}
where $kM_1R_1^2$ is the moment of inertia of the accretor, $\tau_s$ is the tidal synchronization timescale, $\omega = \Omega_{\rm spin} - \Omega_{\rm orbit}$ is the spin asynchronicity, and $r_{\rm h}$ is the dimensionless (scaled by the orbital separation) radius of the orbit around the accretor that holds the same angular momentum as the accreted mass. This radius must be solved for numerically; a convenient fit for $r_{\rm h}$ can be found in \cite{Verbunt:88} (note that their definition of $q$ is the inverse of ours). 

The Roche lobe overfill factor is defined as:
\begin{equation}
    \Delta = R_2-R_L,
\end{equation}
where $R_2$ is the donor star radius and $R_L$ is the radius of the binary Roche lobe. In this work we calculate $R_L$ using the fit in \cite{Eggleton:83}. This overfill factor determines the mass transfer rate and evolves as \cite{Marsh:04}:
\begin{equation} \label{eq: Ddot}
    \dot{\Delta} = \left(R_2\zeta_2-R_L \zeta_{R_L} \right)\frac{\dot{M}_2}{M_2}+\frac{2 \beta R_L}{3 T_{\rm obs} \alpha},
\end{equation}
where
\begin{equation}
    \zeta_2 = \frac{d \log R_2}{d \log M_2},
\end{equation}
and
\begin{equation}
    \zeta_{R_L} = \frac{d \log(R_L/a)}{d \log M_2}.
\end{equation}
We consider steady state mass transfer in this work and assume that the overfill factor is slowly evolving and that orbital angular momentum lost due to mass transfer is returned to the orbit through tidal synchronization. We therefore set $\dot{\Delta} = \omega = r_h = 0$. This approximation yields a slightly more negative value of $\beta$ than the full solution as this feedback mechanism is not perfectly efficient. Our assumptions work reasonably well for short tidal synchronization timescales $\tau_s \leq 50$ yr for an initial mass ratio $q=1/3$.

Using Eq. \ref{eq: Ddot} to get $\dot{M}_2$ and plugging the result back into Eq. \ref{eq: betadef} then grants:
\begin{equation}
    \frac{\beta}{3 T_{\rm obs}\alpha} = -\frac{\dot{J}_{\rm GR}}{J_{\rm orb}}\left[\frac{R_2\zeta_2 - R_L\zeta_{R_L}}{R_2\zeta_2+R_L(2-2q-\zeta_{R_L})} \right].
    \label{eq:beta_theoretical}
\end{equation}
For a binary with initial component masses $M_1 = 0.6 M_{\odot}$ and $M_2 = 0.2 M_{\odot}$ evolved for a short time past Roche lobe overflow to a mass ratio $q=0.332$, $\zeta_{R_L} = 0.351$, $\zeta_2 = -0.354$, and $R_2 = 1.41\times10^7$ m. These values grant an estimate from Eq. \ref{eq:beta_theoretical} of $\beta = -53.9$ while full numerical evolution coupling $\omega$, $\Delta$, and $a$ as detailed in \cite{Marsh:04} grants our fiducial $\beta = -49.5$. For our fiducial value of $\beta$ we use the fit from \cite{Verbunt:88} to get $r_h = 0.136$. We used a tidal synchronization time $\tau_s = 10$ yr, which is roughly the maximum allowed to achieve stable mass transfer within a drop in mass ratio of $0.001$. The result is unchanged when using a shorter synchronization time. On the other hand, a $\tau_s$ between 10 and 100 years leads to oscillations around a value near the one predicted by Eq. \ref{eq:beta_theoretical}, while $\tau_s\gtrsim 10^3\,{\rm yr}$ causes unstable mass transfer. Theoretically, both weakly nonlinear interactions among WD gravity modes \cite{Yu:20a} or strongly nonlinear traveling gravity waves \cite{Fuller:12a, Fuller:13, Burkart:13} can produce $\tau_s\lesssim 10\,{\rm yr}$ for a binary similar to the fiducial one considered here.

In this work we set $\phi_0 = 0$ and specify values of $\gamma$ and $t_b$ used in the generation of each figure.
Our waveform is given by:
\begin{equation}
    h(t) = A \cos(\phi(t)),
\end{equation}
where
\begin{equation}
    \phi(t) = \phi_0 + 2\pi \alpha t + \pi \beta t^2 + 2\pi \gamma \left(t-t_b\right)\Theta(t-t_b).
    \label{eq:phi_vs_t}
\end{equation}
Here $\Theta$ is the Heaviside step function.
We define an inner product in the usual way as:
\begin{equation}
    \langle a|b \rangle = 4\int_0^\infty\frac{\tilde{a}^*(f)\tilde{b}(f)}{S_n(f)}d f \simeq \frac{2}{S_n(f_0)}\int_0^1a(t)b(t)d t,
\end{equation}
where $S_n(f)$ is the power spectral noise density.
The Fisher matrix is then given as
\begin{equation}
    \Gamma_{ij} = \left\langle \frac{\partial h}{\partial\theta^i}\bigg|\frac{\partial h}{\partial\theta^j}\right\rangle \simeq     \frac{2}{S_n(f)}\int_0^{1} d t \frac{\partial h}{\partial\theta^i}\frac{\partial h}{\partial\theta^j},
\end{equation}
where $\theta^i$ are the parameters $(\log A,\phi_0, \alpha, \beta, \gamma, t_b)$. The SNR $\rho$ is given as: 
\begin{equation}
    \rho = \frac{\langle h|h\rangle}{{\rm rms}\langle h|n \rangle} = \langle h|h\rangle^{1/2}.
\end{equation}
We define $\tau_a = \left[1-(t_b)^a\right]$ and $\upsilon_a = (t_b)^a-at_b+a-1$. In the approximation where many GW cycles are measured by LISA, $fT_{\rm{obs}} \gg 1$, the Fisher matrix is analytically given by:

\begin{equation}
  \Gamma \simeq \rho^2\left(
    \begin{array}{cccccc}
      1 & 0 & 0 & 0 & 0 & 0 \\
    0 & 1 & \pi & \frac{\pi}{3} & \pi \upsilon_2 & -2 \pi \gamma\tau_1 \\
    0 & \pi & \frac{4\pi^2}{3} & \frac{\pi^2}{2} & \frac{2\pi^2}{3} \upsilon_3 & -2 \pi^2 \gamma \tau_2 \\
    0 & \frac{\pi}{3} & \frac{\pi^2}{2} & \frac{\pi^2}{5} & \frac{\pi^2}{6} \upsilon_4 & -\frac{2}{3} \pi^2 \gamma \tau_3 \\
    0 & \pi \upsilon_2 & \frac{2\pi^2}{3} \upsilon_3 & \frac{\pi^2}{6} \upsilon_4 & \frac{4\pi^2}{3} \left(1-t_b\right)^3 & -2\pi^2\gamma \upsilon_2\\
    0 & -2 \pi \gamma \tau_1 & -2 \pi^2 \gamma \tau_2 & -\frac{2}{3} \pi^2 \gamma \tau_3 & -2\pi^2\gamma \upsilon_2 & 4\pi^2\gamma^2 \tau_1
    \end{array}
  \right)
  \label{eq:Fisher}
\end{equation}

In this work we make use of a Reversible Jump Markov Chain Monte Carlo algorithm (RJMCMC) \cite{Geyer:94, Green:95} to jump between models with and without the effects of a burst. Specific details of our trans dimensional jump probabilities are provided in \ref{adx: RJMCMC details}. 
When computing jumps we make use of the augmented Fisher matrix:
\begin{equation}
    \Upsilon_{ij} = \Gamma_{ij} - \partial_i\partial_j\ln p(\theta|\mathcal{H}),
\end{equation}
where $p(\theta|\mathcal{H})$ is the prior on our waveform parameters. We use a symmetric in sign, uniform in log prior for $\gamma$ where $|\Delta f|/f = |\gamma|/\alpha \in [10^{-8}, 10^{-3}]$ and a uniform prior range for $t_b$ of $[0,1]$. 
For our non-burst parameters, we use uniform priors of [$1,10^3$], [$-\pi,\pi$], and [$-T_{\rm obs}^2,T_{\rm obs}^2$] for $A$, $\phi_0$, and $\beta$. For $\alpha$ we use a prior of $\pm 0.5$ mHz around the true frequency $f$ which corresponds to $\pm6.31\times10^4$ around the injected value of $\alpha$ for a four year observation time. These priors on the non-burst parameters have no impact on the detectability of the burst as long as they are not too restrictive; detectability only depends on the change in the prior space when switching models, as described in the next section.

\section{Detection Thresholds} \label{sec:Detection Thresholds}
Determining the SNR required for a confident detection of a burst event through use of a RJMCMC can be prohibitively computationally expensive. Therefore, we seek to cheaply estimate the required SNR for given burst parameters without the need for a full RJMCMC run.
In Bayesian statistics, the probability for one model hypothesis $\mathcal{H}$ over another is given by the odds ratio:
\begin{equation}
    \frac{p(\mathcal{H}_2|\bf{d})}{p(\mathcal{H}_1|\bf{d})} = \frac{p(\bf{d}|\mathcal{H}_2)p(\mathcal{H}_2)}{p(\bf{d}|\mathcal{H}_1)p(\mathcal{H}_1)} = \mathcal{B}_{21}\mathcal{P}_{21},
\end{equation}
where $\mathcal{B}_{21}$ and $\mathcal{P}_{21}$ are the Bayes factor and prior odds ratio for model 2 over 1, respectively. From this point forward we take $\mathcal{H}_2$ to be our burst model while $\mathcal{H}_1$ is our non-burst model. We can estimate the Bayes factor using the Laplace approximation to the evidence, which assumes that the region surrounding the maximum of the posterior is well approximated by a multivariate Gaussian
\cite{Azevedo-Filho:94, Cornish:19}:
\begin{equation}
    p(d|\mathcal{H}) \approx p(d|\vec{\theta}, \mathcal{H})|_{\rm{MAP}}\left(\frac{p(\vec{\theta}|\mathcal{H})|_{\rm{MAP}}}{\sqrt{\rm{det}(\Upsilon/2\pi)}}\right),
\end{equation}
where MAP stands for \emph{maximum a posteriori}. For the simple problem considered in this work we take the MAP parameters to be our injected values when performing parameter estimation in the burst model. The term in the parentheses is commonly known as the ``Occam Factor,'' hereafter denoted by $\mathcal{O}$, and is the penalty imposed by the MCMC algorithm for the complexity of the model. When uniform priors are used, the Occam factor reduces to the posterior volume over the prior volume.

Next we define the fitting factor: the measure of how well we can recover a signal from one template family using a different family. We start by defining the waveform match:
\begin{equation} \label{eq:match definition}
    M = \frac{\langle h_1|h_2\rangle}{\sqrt{\langle h_1|h_1\rangle\langle h_2|h_2\rangle}}.
\end{equation}
The fitting factor $FF$ is then defined as the maximum match that can be achieved by varying the parameters in template family 1. In our model the fitting factor is the maximum match obtained by varying our non-burst parameters to recover a given burst waveform. In this work we assume that the amplitudes of the templates are the same so that $\langle h_1|h_1\rangle = \langle h_2|h_2\rangle= \rho^2$. To obtain the fitting factor we shift each parameter in our non-burst model by some amount $\mu^i_{\rm tr} \rightarrow \mu^i_{\rm tr} + \Delta \mu^i = \mu^i_{\rm bf}$, where $\mu^i_{\rm bf}$ are the best fit parameter values recovered when using templates in our non-burst model. By definition, the shift in our burst amplitude is given by $\Delta \gamma = -\gamma$. With this defined value the best fit waveform is given by:
\begin{equation}
    h(\mu_{\rm bf}) = A \cos\left[\phi_0 +\Delta\phi_0 + 2\pi(\alpha + \Delta \alpha)t + \pi(\beta + \Delta \beta)t^2\right],
\end{equation}
which has no dependence on $\Delta t_b$. However, the series expansion of the match that we examine in the following discussion does depend on $\Delta t_b$ at each order. We therefore make the choice $\Delta t_b = 0$ to make the expansion more accurate.

We assume that the parameter shifts are small and expand the match around $\Delta \theta^i = 0$. We ignore terms with an unequal number of derivatives on each side of the inner product, such as $\langle h|h,_i\rangle$. These terms are only non-vanishing in the many cycles limit $f T_{\rm{obs}} \gg 1$ for i = 0 and we assume $\Delta A = 0$ which kills these terms. Under these assumptions the match can be expanded around $\Delta\theta^i=0$:
\begin{equation} \label{eq:match expansion}
    M = 1 - \frac{1}{2}\frac{\Gamma_{ij}}{\rho^2} \Delta\theta^i \Delta \theta^j+....
\end{equation}
We split our shifts into known and unknown values: $\Delta \theta^i = \{\mu^i, -\gamma, 0\}$. We then find the unknown shifts through the expansion for the match, their locations are defined through:
\begin{equation} \label{eq:match derivative}
    \frac{\partial M}{\partial \Delta \mu^i} = -\frac{1}{2\rho^2}\left[(\Gamma_{\mathcal{H}_1})_{ij}\Delta \mu^j - \gamma\Gamma_{i \gamma}\right] = 0,
\end{equation}
where $\Gamma_{\mathcal{H}_1}$ is the Fisher matrix of our non burst parameters and is the leading principal minor of order four of Eq. \ref{eq:Fisher}. 
We can then solve for the shifts of our non-burst parameters:
\begin{equation} \label{eq:parameter shifts}
    \Delta \mu^j = \gamma\Gamma_{i\gamma}(\Gamma_{\mathcal{H}_1}^{-1})^{ij}
\end{equation}
In the simple case examined here it is also possible to obtain $\mu^i_{\rm bf}$ by directly fitting the injected phase with the non-burst parameters. 
We can then calculate the fitting factor from the match as:
\begin{equation} \label{eq:Fitting factor}
    FF = 
    \frac{1}{\rho^2}\left\langle h_1(\mu^i_{\rm bf})|h_2(\theta^i_{\rm tr})\right\rangle
\end{equation}

\begin{figure}
    \centering
    \includegraphics[width=0.8\linewidth]{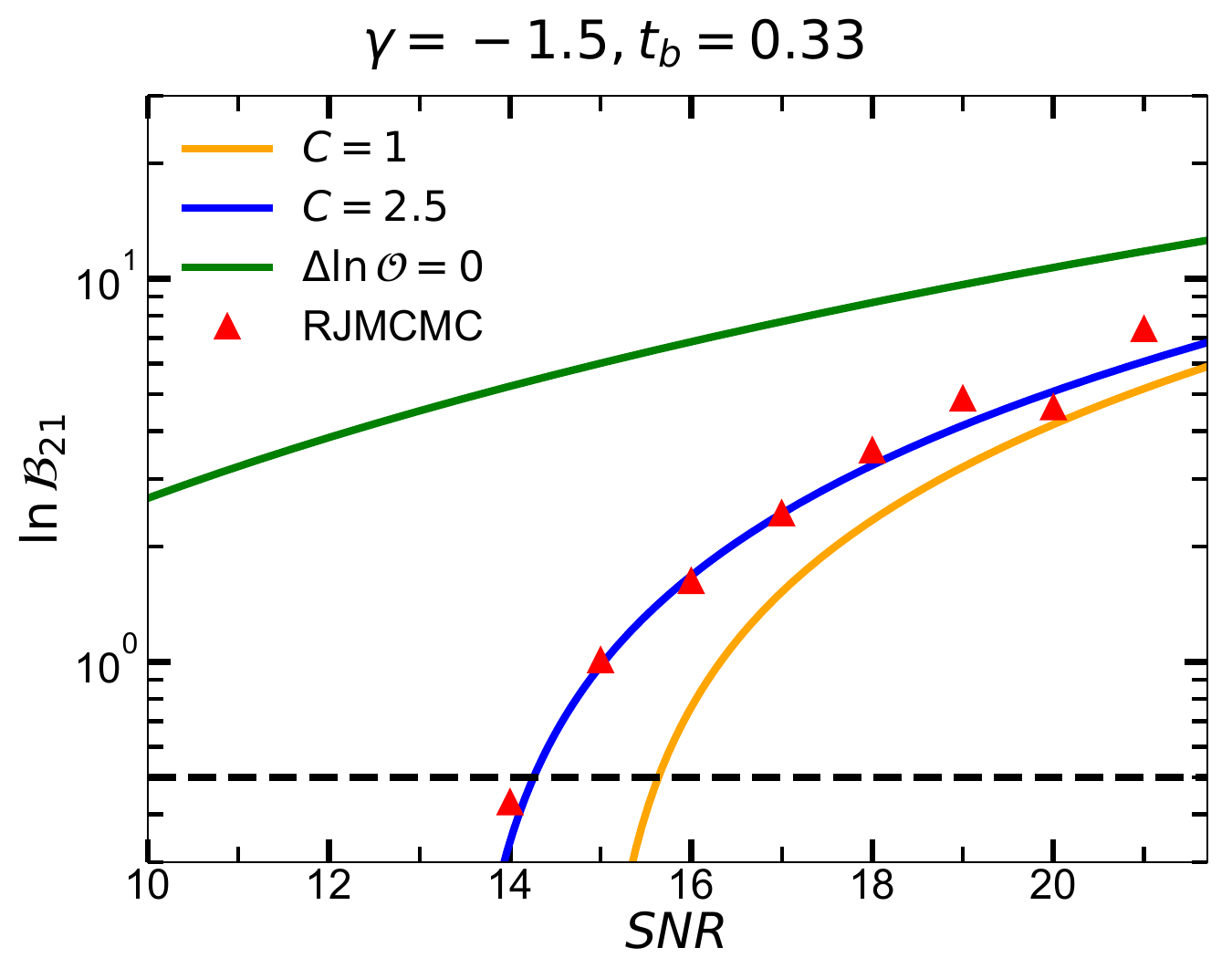}
    \caption{Results for Eq. \ref{eq:logBF expansion} with and without the Occam factor. We compare these results to Bayes factors obtained from our RJMCMC through computation of the odds ratio. The black dashed line shows $\ln \mathcal{B}_{21} = 0.5$, the value used when generating both panels of Fig. \ref{fig:Detection SNRs}.
    }
    \label{fig:lnBFs}
\end{figure}

With these ingredients, we can approximate the natural log of the Bayes factor as \cite{Cornish:11}:
\begin{equation} \label{eq:logBF expansion}
    \ln \mathcal{B}_{21} = (1-FF^2)\frac{\rho^2}{2} + \Delta \ln \mathcal{O}.
\end{equation}
When the Occam factor is ignored, one can estimate the threshold SNR at which two waveforms differing by a shift in parameters, $\Delta \theta^i$, can be distinguished.
Taking the expansion for the match from Eq. \ref{eq:match expansion} and replacing $FF$ with the match in Eq. \ref{eq:logBF expansion} while neglecting the Occam factor, setting $\ln \mathcal{B}_{21} = 1/2$, and taking $FF$ to be near unity grants the condition $\Gamma_{ij}\Delta\theta^i\Delta\theta^j = 1$. 
Assuming that $\Delta\gamma = -\gamma$ and that the shifts of all other parameters are given by Eq. \ref{eq:parameter shifts}, we get $|\gamma| = \sqrt{\left(\Gamma^{-1}\right)^{\gamma \gamma}}$. 
This gives a ``Fisher'' detection SNR of:
\begin{equation}
     \rho_F =\frac{\sqrt{3}}{\pi |\gamma|}\sqrt{\frac{1-4t_b(1-t_b)\left[2-5t_b(1-t_b)\right]}{t_b^3(1-t_b)^3\left\{1-5t_b(1-t_b)[1-t_b(1-t_b)]\right\}}}.
\end{equation}
This equation gives an approximation to the intersection of the green and black lines in Fig. \ref{fig:lnBFs}.
For simplicity in solving Eq. \ref{eq:logBF expansion}, we use the standard Fisher matrix $\Gamma$ in place of the augmented Fisher matrix $\Upsilon$ when computing the Occam factor. These matrices converge in the high SNR limit so this replacement has little impact on accuracy when computing a threshold SNR. 

Using the definition of the Occam factor, we have
    \begin{eqnarray}[rl]
\Delta \ln \mathcal{O} & \simeq \left[ p(\gamma, t_b| \mathcal{H}_2)\frac{\sqrt{{\rm{det}(\Gamma}_{\mathcal{H}_1}/2\pi)}}{\sqrt{{\rm{det}(\Gamma}_{\mathcal{H}_2}/2\pi)}} \right] \\
    &=\ln\left\{2\pi \left[ \frac{1}{\gamma}p(\ln\gamma| \mathcal{H}_2)p(t_b| \mathcal{H}_2)\right]\frac{\sqrt{{\rm{det}(\Gamma}_{\mathcal{H}_1})}}{\sqrt{{\rm{det}(\Gamma}_{\mathcal{H}_2})}}\right\},
    \end{eqnarray}
where the expression is evaluated at the MAP. In the second equality, we have used the fact that the priors on $\gamma$ and $t_b$ are uncorrelated, and changed the probability of $\gamma$ to that of $\ln \gamma$.
Since the priors are uniform in $\ln \gamma$ and $t_b$, we have $p(\ln\gamma| \mathcal{H}_2)p(t_b| \mathcal{H}_2)=1/V_b$, with 
\begin{equation}
    V_b=2\int_{\ln (10^{-8}\alpha)}^{\ln (10^{-3}\alpha)} d\ln\gamma\int_0^1 \, dt_b,
\end{equation}
where the factor of 2 accounts for both signs allowed by $\gamma$. 
This leads to
\begin{equation} \label{eq:occam}
    \Delta \ln \mathcal{O} \simeq \ln(\mathcal{O}_0) + \ln C - 2 \ln \rho,
\end{equation}
where the SNR independent component of the Occam factor is:
\begin{equation}
    \mathcal{O}_0 = \frac{\sqrt{3}}{\pi t_b^2(1-t_b)^2{\gamma}^2 V_b}  (1-5t_b + 10t_b^2 - 10t_b^3+5t_b^4)^{-1/2},
\end{equation}
and $C$ is a fudge factor introduced to account for the multi-modal nature of the burst posterior that
is not captured by the Fisher matrix formulism. 

The expansion given in Eq. \ref{eq:logBF expansion} assumes that the posterior can be approximated by a multivariate gaussian. We find that this assumption is not entirely valid in the case of a nova burst as the maximum match achieved in the non-burst model can be equaled or exceeded by jumping to the burst model with a small burst or with a time of burst centered between the injected time of burst and the edge of the observation period that is further from the injected time.

\begin{figure}
    \centering
    \includegraphics[width=0.9\linewidth]{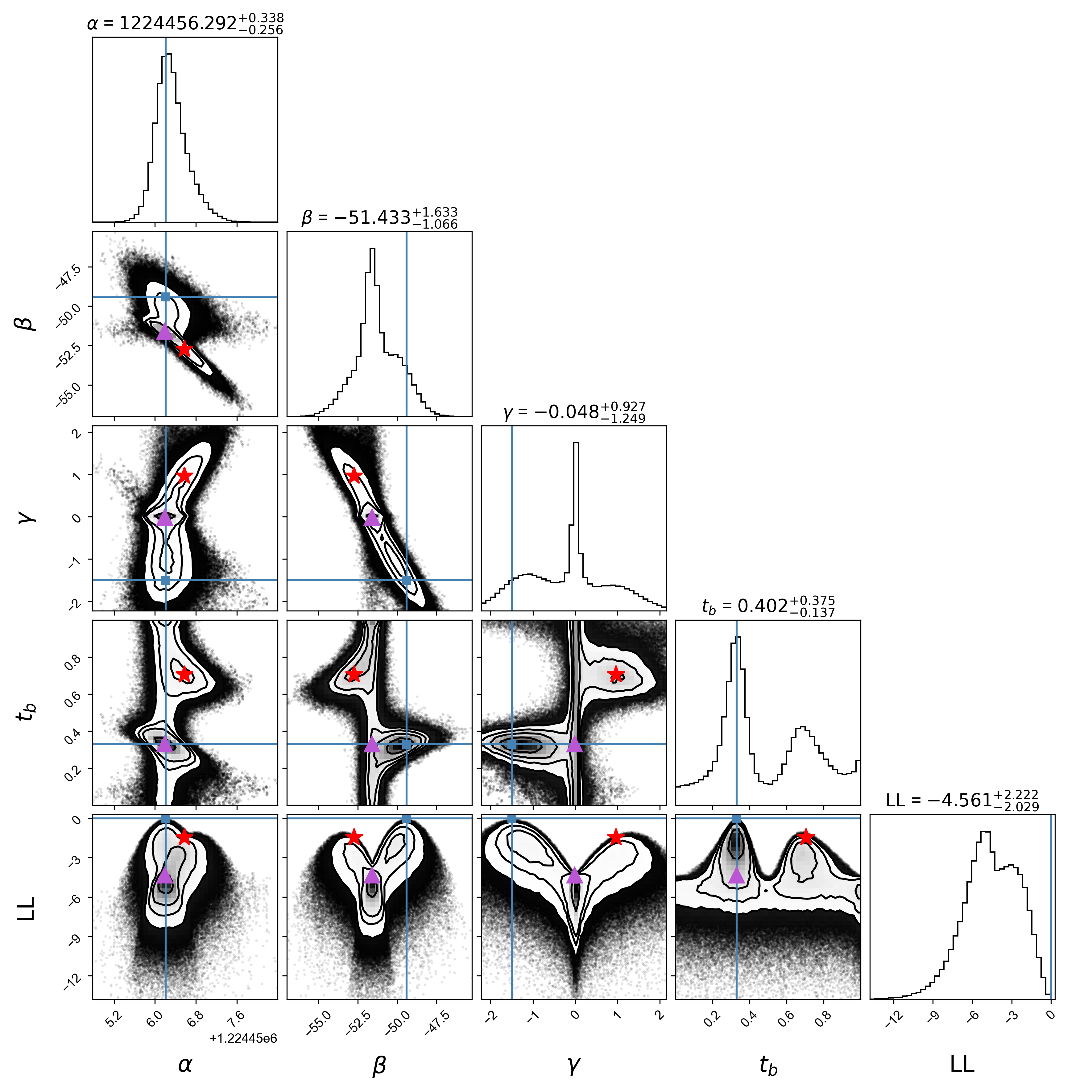}
    \caption{Parameter estimation results for an injected waveform with $\gamma=-1.5$ and $t_b=0.33$. The blue lines show injected parameters. The purple triangles correspond to an optimal set of parameters where $\gamma$ is much smaller than the injected value while 
    the red stars correspond to optimal parameters for a nova shift with opposite sign to the true shift. Here LL is the log likelihood for each point.}
    \label{fig:Likelihood Corner}
\end{figure}

\begin{figure}
    \centering
    \includegraphics[width=0.85\linewidth]{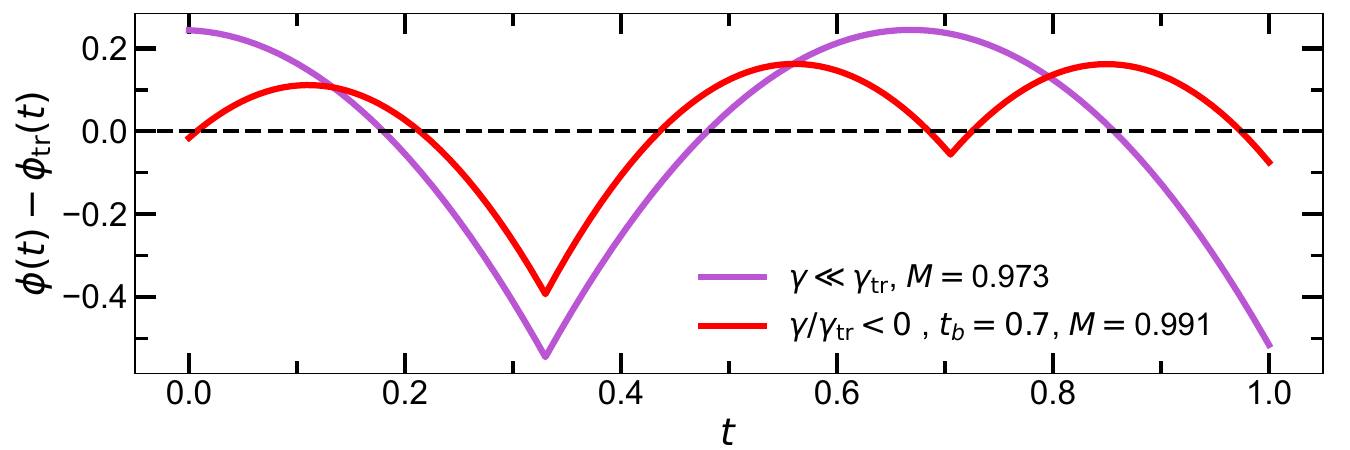}
    \caption{Here we show the deviation from the true phase for the two sets of parameters highlighted in Fig. \ref{fig:Likelihood Corner} using the same colors.}
    \label{fig:Phase differences}
\end{figure}

We showcase these features using parameter estimation in Fig. \ref{fig:Likelihood Corner}. We filter our results and only display points in the burst model. The purple triangles show a set of parameters with optimal match for $|\gamma| \simeq 0$ while the red stars show a set of parameters near optimal match for $\gamma$ of the opposite sign of the injected value. We show the deviation from the true phase for these sets of parameters as a function of time in Fig. \ref{fig:Phase differences}. These plots show the multi-modal nature of the posterior that results in a $\sim 2-3$ times increase in burst model posterior volume. This results in a higher Bayes factor than approximated in Eq. \ref{eq:logBF expansion}. We model this increase using the factor $C$. See \ref{adx:Posterior} for more details on how $C$ represents the multi-modality of the posterior.

\begin{figure}
    \includegraphics[width=\linewidth]{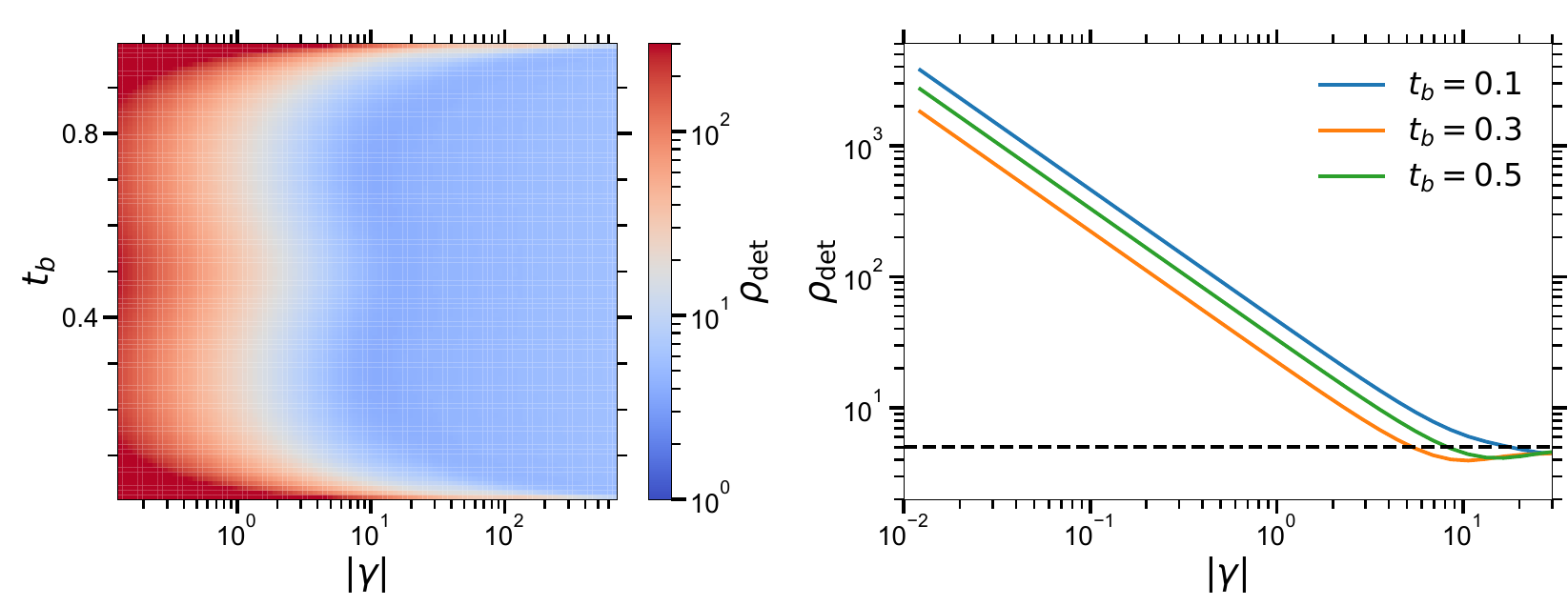}
    \caption{Solutions to Eq. \ref{eq:rho_det} for selected values of $\gamma$ and $t_b$ with $\ln \mathcal{B}_{21} = 0.5$ and $C=1$. The left panel shows solutions for a binary at 9.7 mHz. The color scale was capped at 1 and 300 to maintain clarity. The shape seen in the left panel is due to degeneracy between $\beta$ and $\gamma$, as considered in Fig. \ref{fig:Phase differences 2}. The right panel shows the full range of SNRs across a wider range of $\gamma$ for three chosen values of $t_b$. The black dashed line shows a SNR of 5 for clarity.}
    \label{fig:Detection SNRs}
\end{figure}

\begin{figure}
    \centering
    \includegraphics[width=0.85\linewidth]{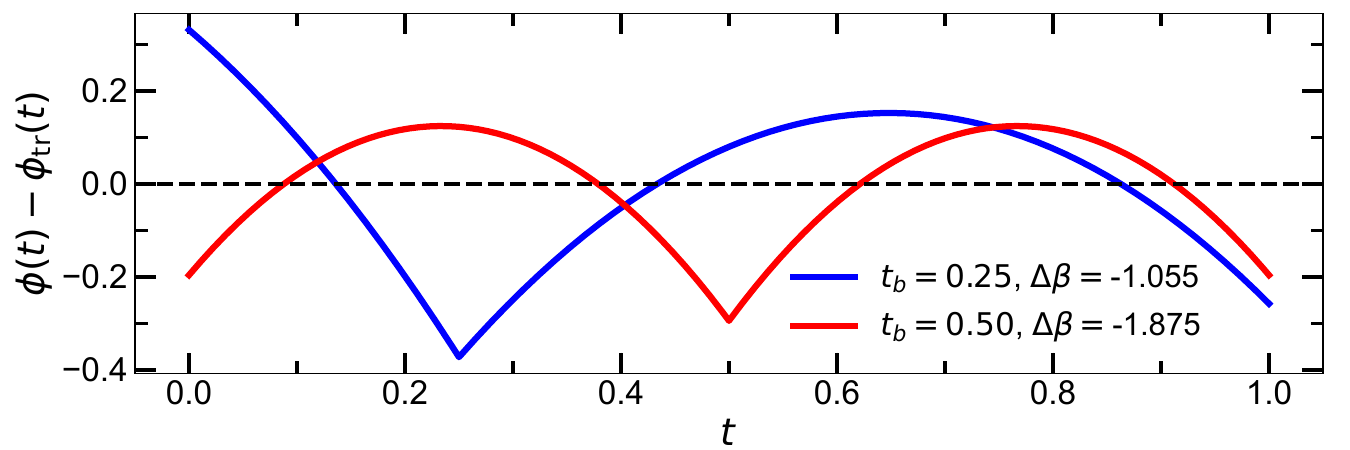}
    \caption{Deviation from the injected phase for best fit non-burst waveforms for burst magnitudes $\gamma = -1$ at times $t_b = 0.25, 0.5$. These show the cause of the $\mathcal{E}$-shape seen in the left panel of Fig. \ref{fig:Detection SNRs}: the average phase deviation is smaller for bursts in the middle of the observation time than at $t_b\sim0.3$ or $t_b\sim0.7$, resulting in an improved fitting factor at the expense of a larger bias in $\beta$. The blue and red lines have fitting factors of 0.987 and 0.994, respectively.
    }
    \label{fig:Phase differences 2}
\end{figure}

\begin{figure}
    \centering
    \includegraphics[width=0.9\linewidth]{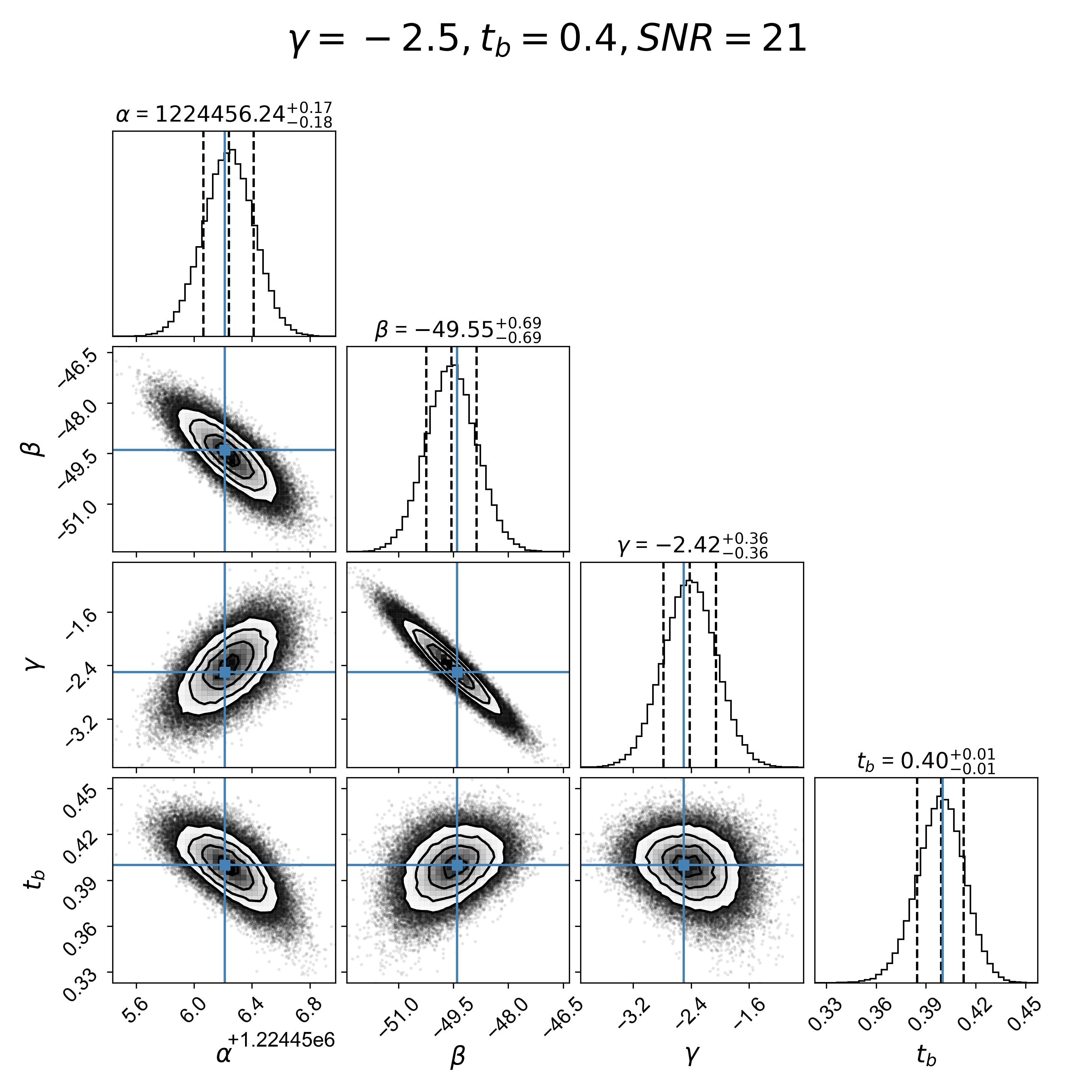}
    \caption{Corner plot showing the parameter estimation results for an injected waveform with $\gamma=-2.5$ and a burst at 1.6 years. The blue lines show the injected parameter values. We use a noise free likelihood calculation so the distributions are expected to center on the true values.
    }
    \label{fig:Detected Corner}
\end{figure}

We then get a transcendental equation for the SNR from Eqs. \ref{eq:logBF expansion} and \ref{eq:occam}:
\begin{equation}
    \rho^2\exp\left(-(1-FF^2)\frac{\rho^2}{2} \right) = \frac{\mathcal{O}_0 C}{\mathcal{B}_{21}}.
\end{equation}
This equation can be solved for the SNR for a chosen Bayes factor via use of the real valued Lambert W function, also known as the product logarithm; useful properties are summarized in \cite{Veberi:12}. We choose the -1 branch to align with our goal of a burst detection SNR. This grants:
\begin{equation} \label{eq:rho_det}
    \rho_{\rm{det}} = \sqrt{\frac{-2}{1-FF^2}W_{-1}\left(-\frac{1}{2}(1-FF^2)\frac{\mathcal{O}_0}{\mathcal{B}_{21}}C \right)}.
\end{equation}
Solutions to Eq. \ref{eq:logBF expansion} with and without the additional Occam factor term are compared with results from our RJMCMC in Fig. \ref{fig:lnBFs}. We use a variable prior ratio which allows us to determine higher Bayes factors efficiently. This method leads to low probabilities for the RJMCMC to make jumps between models, causing slightly noisy results at higher SNRs.
Values of $\rho_{\rm{det}}$ for $\ln\mathcal{B}_{21} = 1/2$ and $C=1$ are presented in Fig. \ref{fig:Detection SNRs}. We show this choice for $\ln\mathcal{B}_{21}$ with a black dashed line in Fig. \ref{fig:lnBFs}. 
Note that the $\mathcal{E}$-shape of the results is due to the correlation between $\beta$ and $\gamma$. 
A burst closer to $t_b=0.5$ can be better compensated by a larger bias in $\beta$.
This causes a higher fitting factor when the burst is near the center of the observation time than for bursts near the first or third quarter points. This detail is shown in Fig. \ref{fig:Phase differences 2}.
We show an example of parameter estimation for a system with an SNR much greater than the detection SNR in Fig. \ref{fig:Detected Corner}

We note that the curves for $\rho_{\det}$ vs $\gamma$ follow the same contours independent of mass ratio. This is due to the fact that the Fisher matrix is independent of $\alpha$ and $\beta$ so changing these parameters does not change the SNR where $\sqrt{(\Gamma^{-1})_{\gamma \gamma}} =  \gamma$, thus not affecting the detection SNR. 
This fact can also be seen when considering the match as a different frequency does not change the relationship between $\alpha$ and $\gamma$ as both scale linearly with frequency for a given fractional shift $\Delta f/f$. Therefore, a change in frequency has no effect on the waveform match. The only effect that different frequencies have on the analysis is the prior range on $\gamma$ depends on frequency as $|\gamma| \in [\alpha\times10^{-8},\alpha\times10^{-3}]$.

\section{Stealth Bias}  \label{sec:Stealth Bias}
\begin{figure}
    \includegraphics[width=\linewidth]{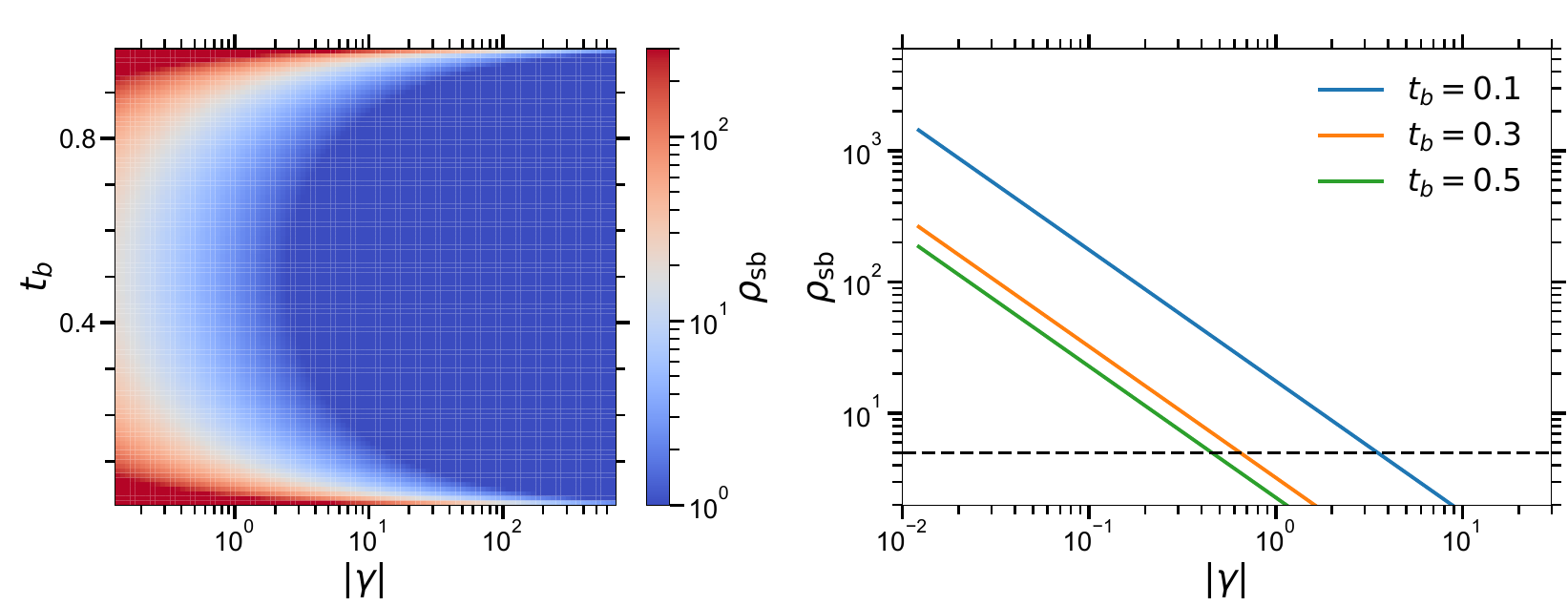}
    \caption{Solutions to Eq. \ref{eq:bias SNR} for selected values of $\gamma$ and $t_b$. The left panel shows solutions for a binary at 9.7 mHz. The color scale was capped at 1 and 300 to maintain clarity. The right panel shows the full range of SNRs across a wider range of $\gamma$ for three chosen values of $t_b$. The black dashed line shows a SNR of 5.}
    \label{fig:Bias SNRs}
\end{figure}

Future GW signals that include the effects of a nova burst can often be well recovered by performing parameter estimation with waveforms excluding the effects of the burst.
This leads to a ``stealth bias'' where if the burst is not explicitly searched for the bias may never be discovered \cite{Vallisneri:13, Vitale:14, Fiacco:24}. For small bursts these biases often persist even when novae are searched for as the SNR required for bias is often much lower than the SNR necessary for burst detection. We can define the parameter space of this stealth bias as the space where the theoretical mismodeling from excluding the effects of the burst causes a bias greater than one expected standard deviation from the noise. We focus this analysis on $\beta$ as this is our only intrinsic binary parameter in the non-burst model, containing information on the binary component masses and donor stellar structure (see Eq. \ref{eq:beta_theoretical}).
To do this we use Eq. \ref{eq:parameter shifts} to calculate the bias in $\beta$ due to theoretical mismodeling:
\begin{equation} \label{eq:delta beta theory}
    \Delta \beta_{\rm th} = 30\gamma t_b^2(1-t_b)^2.
\end{equation}
We then invert $\Gamma_{\mathcal{H}_1}$ to find the expected spread in $\beta$ due to noise when performing parameter estimation in the non-burst model:
\begin{equation} \label{eq:delta beta noise}
    \Delta \beta_{\rm n} = \sqrt{(\Gamma_{\mathcal{H}_1}^{-1})_{\beta \beta}} = \frac{6\sqrt{5}}{\pi \rho}.
\end{equation}
These equations then grant a stealth bias threshold SNR where $\Delta \beta_{\rm{th}} = \Delta\beta_{\rm n}$:
\begin{equation} \label{eq:bias SNR}
    \frac{1}{\rho_{\rm sb}} = \pi \sqrt{5}|\gamma| t_b^2(1-t_b)^2.
\end{equation}
Values of $\rho_{\rm sb}$ for selected values of $\gamma$ and $t_b$ are provided in Fig. \ref{fig:Bias SNRs}.

\begin{figure}
    \centering
    \includegraphics[width=0.7\linewidth]{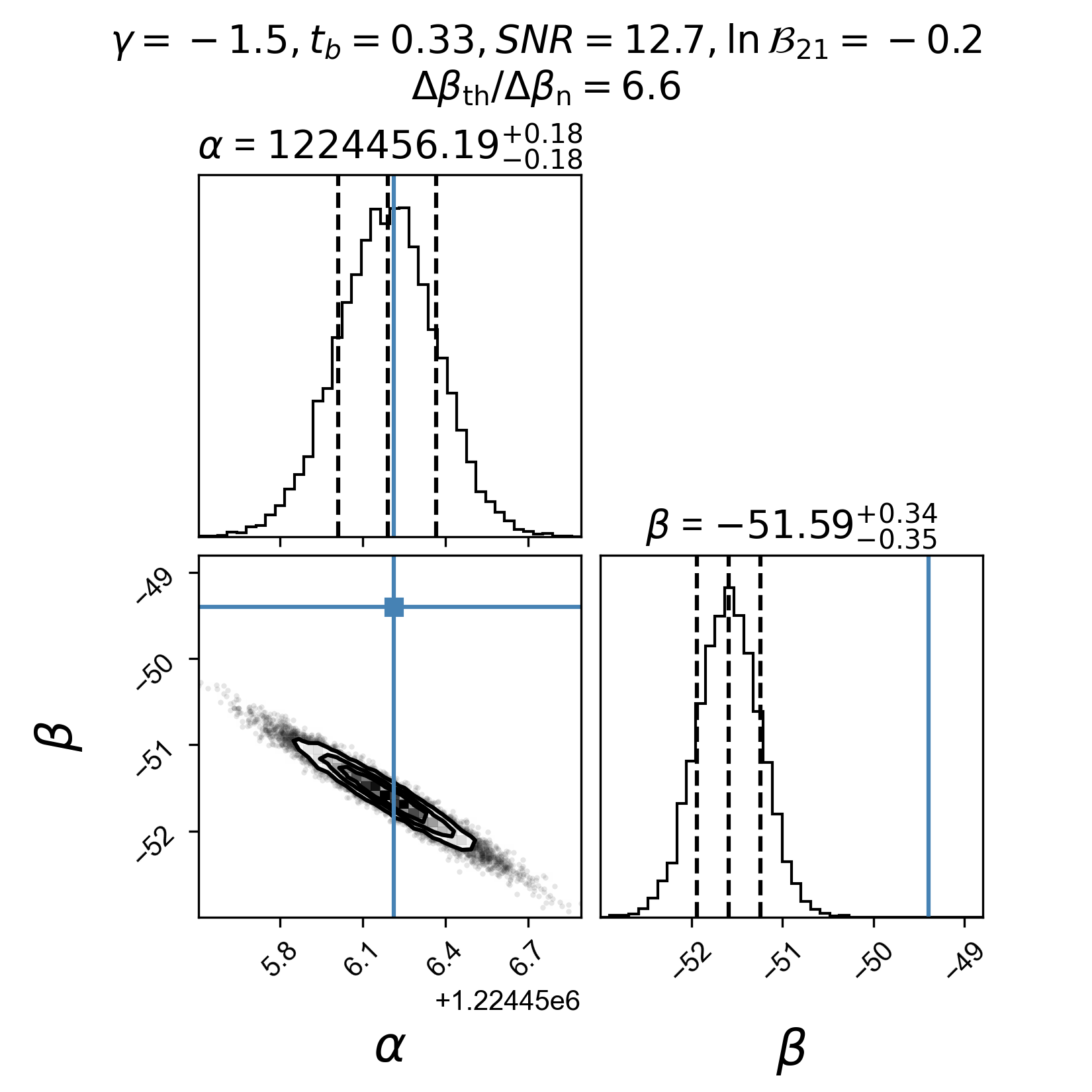}
    \caption{Corner plot showing the parameter estimation results for $\alpha$ and $\beta$ from a non-burst waveform for an injection with $\gamma=-1.5$ and a burst at 1.32 years. We find a Bayes factor of 0.82 using our RJMCMC. Despite this Bayes factor less than 1, the result in $\beta$ is biased by 6.6 times the Fisher matrix noise spread. The blue lines show the injected parameter values.}
    \label{fig:Bias SNR Corner}
\end{figure}

Comparing Fig. \ref{fig:Bias SNRs} to Fig. \ref{fig:Detection SNRs} confirms that in many cases the detection SNR is significantly higher than the stealth bias threshold SNR for a given pair of burst parameters. In some cases this leads to a mismodeling bias much greater than statistical uncertainty even while the Bayes factor prefers a model without a burst.
This bias can be written as the ratio of the theoretical to expected noise biases:
\begin{equation} \label{eq:bias ratio beta}
    \left|\frac{\Delta \beta_{\rm th}}{\Delta \beta_{\rm n}} \right| = \rho\pi \sqrt{5} |\gamma|t_b^2(1-t_b)^2.
\end{equation}
An example with a bias in $\beta$ of $\Delta \beta_{\rm th} = 6.6\Delta\beta_n$ with a Bayes factor of 0.82 is shown in Fig. \ref{fig:Bias SNR Corner}.
Similarly, the theoretical and expected noise biases for $\alpha$ are given by:
\begin{equation}
    \Delta \alpha_{\rm th} = -\gamma(1-t_b)^2(3t_b-1)(5t_b+1),
\end{equation}
\begin{equation}
    \Delta \alpha_{\rm n} = \frac{4\sqrt{3}}{\pi\rho}.
\end{equation}
These then grant a ratio of biases for $\alpha$ given by:
\begin{equation} \label{eq:bias ratio alpha 1}
    \left|\frac{\Delta \alpha_{\rm th}}{\Delta \alpha_{\rm n}} \right| =\frac{\rho\pi|\gamma|}{4\sqrt{3}}(1-t_b)^2|3t_b-1|(5t_b+1).
\end{equation}
Note that in calculating $\Delta \alpha_{\rm n}$ and $\Delta \beta_{\rm n}$, we use the Fisher estimates for the statistical uncertainty calculated from $\Gamma_{\mathcal{H}_1}$. The actual posterior distribution is slightly wider than this optimistic error calculation.
For example, the posterior scaled shift in $\beta$ for the posterior shown in Fig. \ref{fig:Bias SNR Corner} is 6.3 times one half the distance between the 16th and 84th percentiles of the recovered posterior.  In cases with low fitting factors, there is additional broadening due to the non burst model being insufficient to match the injection.

\section{Model Agnostic Tests} \label{sec:stealth bias 2}

In this work, we propose a simplified model for nova bursts that only adds two new parameters to include in parameter estimation on future LISA data. Here we propose a model-agnostic method to search for bursts and similar phenomena that does not rely on the specific form of GW phase given by Eq. \ref{eq:phi_vs_t}.
We consider performing parameter estimation on two equal fractions of the total observation time and comparing results to search for bursts. As shown in the previous section, many bursts will bias parameter estimation results far greater than is expected from the noise. Future parameter estimation results can be compared between two equal time observation periods to propose nova candidate systems for thorough followup analysis.

Theoretical modeling of classical novae has shown that only outbursts for high mass primary stars in binary systems with very high accretion rates are expected to have multiple nova outbursts within the LISA mission time. Furthermore, these bursts will be much smaller than average and will likely have a minimal effect on the GW waveform \cite{Yaron:05}. Therefore, we focus on cases where one half of the LISA observation period includes the effects of a burst while the other does not.

When comparing values of $\alpha$ for two consecutive two-year observation periods, we need to add the values of $\alpha$ and $\beta$ from the first period to compare to $\alpha$ for the second period. When the burst occurs in the second two year observation period, we can use Eq. \ref{eq:bias ratio alpha 1} as the values for $\alpha$ and $\beta$ recovered from the first period will be unbiased. However, when the burst occurs in the first two-year observation time $\Delta \alpha_{\rm th,2}=\gamma$ and $\Delta \beta_{\rm th,2}=0$. Therefore the equivalent ratio to Eq. \ref{eq:bias ratio alpha 1} is given as:
\begin{equation} \label{eq:bias ratio alpha 2}
    \left|\frac{\Delta \alpha_{\rm th,1} + \Delta\beta_{\rm th,1}-\gamma}{\Delta \alpha_{\rm n}} \right| =\frac{\rho\pi|\gamma|}{4\sqrt{3}}t_b^2(6-5t_b)|2-3t_b|.
\end{equation}
The difference between the two recovered values for $\beta$ is still given by Eq. \ref{eq:bias ratio beta} for a burst in either observation period as the observation period without a burst will be unbiased. 

\begin{figure}
\includegraphics[width=\linewidth]{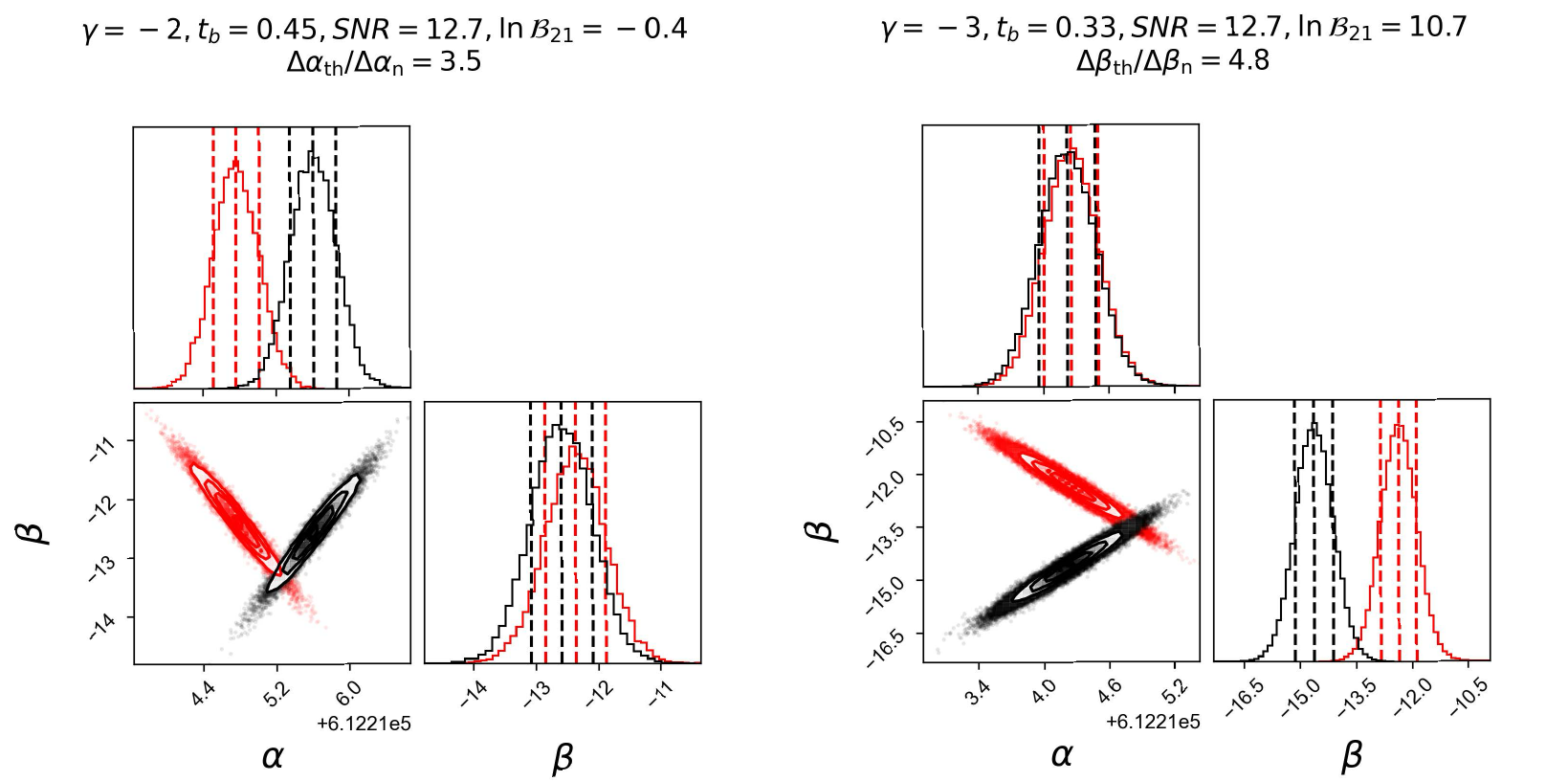}
\caption{
    Parameter estimation for two consecutive two-year observation periods for two different injected burst waveforms recovered without a burst. The red distribution shows parameter estimation results for the second observation period. The black distribution shows the parameter estimation results for the first observation period where the y-axis shows $\beta$ and the x-axis shows $\alpha +\beta$ for accurate comparison with the red distribution. For both plots the titles show the values of the parameters, SNRs, and Bayes factors for the full four-year observation time. 
    }
    \label{fig:Bias Portions}
\end{figure}

We showcase two examples of biases in $\alpha$ and $\beta$ in Fig. \ref{fig:Bias Portions}. 
The titles of the two panels in Fig. \ref{fig:Bias Portions} show the parameter values, SNRs, and Bayes factors for a full four year observation time. 
However, when computing the shifts in $\alpha$ and $\beta$
the values for $\gamma$, $t_b$, and the SNR should be those for the two year observation time with the burst.
The ratio of noise spreads $\Delta\beta_n/\Delta\alpha_n=\sqrt{15}/2 \approx 1.94$ is independent of SNR and is roughly equal to two. Parameter estimation results for $\alpha$ and $\beta$ are highly correlated and adding $\alpha$ and $\beta$ results in a flip in correlation but has little effect on the total spread in $\alpha$. This can be seen in Fig. \ref{fig:Bias Portions} as the black and red distributions have similar size spreads in $\alpha$ but have opposite correlation between $\alpha$ and $\beta$.

\section{Conclusion} \label{sec:Discussion}
The first main result of this paper is that we can reliably measure the magnitude and direction of frequency shifts from classical nova-like bursts in DWDs at SNRs above 15 for $|\gamma| = 1.5$ which corresponds to a fractional frequency shift of $|\Delta f|/f\sim 10^{-6}$ for a fiducial frequency of 9.7 mHz, as shown in Figs. \ref{fig:lnBFs} and \ref{fig:Detection SNRs}. 
The second result is that these bursts, if unaccounted for, can significantly bias parameter estimation results, especially the value of $\beta\propto \dot{f}$ for systems with SNRs of 5-15 (Fig. \ref{fig:Bias SNRs}). This bias often persists when the Bayes Factor for the burst model over the non-burst model is less than unity. This will result in a biased recovery of the frequency evolution, leading to errors in the inferred component masses, donor WD structure, and physics of mass transfer (Eq. \ref{eq:beta_theoretical}).
We find that the SNRs for detection and bias in $\beta$ for a specific $\gamma$ and $t_b$ are independent of the binary frequency and first frequency derivative. We also find that these SNRs can be predicted in a computationally inexpensive way without the need for a RJMCMC. Finally, we show that bursts may be identified in a model agnostic way by comparing two halves of future LISA data as demonstrated for a noise-free case in Fig. \ref{fig:Bias Portions}. This model agnostic method may be useful for identifying other, similar phenomena such as mode resonances \cite{Hansen:75, Rathore_2005, Flanagan_2007, Yu_2016, Lau_2021} or bursts in neutron star - WD binaries \cite{Galloway:21, Kormpakis:25}.

We make several simplifying assumptions about the waveforms studied. First, we only consider waveforms with one burst during the LISA observation time and assume that these bursts affect the orbital frequency instantaneously. These assumptions are motivated by results from \cite{Yaron:05} which show that in the majority of the parameter space in $M_1$ and $\dot{M}_1$ the time of mass loss is much less than the expected LISA observation period and the recurrence period is longer than the observation time. However, these assumptions are somewhat contradictory as more rapid bursts have shorter recurrence periods and longer recurrence times lead to more prolonged mass loss. Future analysis for a small number of systems may need to relax one of these assumptions and allow for a prolonged period of mass loss or multiple instantaneous bursts. These scenarios are both relatively unlikely as systems with multiple bursts will have minimal effect on the orbital frequency and more prolonged, larger bursts are far less likely as their recurrence periods are much longer.

Relaxing either of the two prior assumptions adds computational complexity to analyzing LISA data. This is also true when performing a search for burst effects under our simple model. In a future work we will carefully analyze the parameter space in $f$ and $\dot{f}$ where systems are unambiguously mass transferring. Systems with $\dot{f} < 0$ are certainly in this regime but not all mass transferring systems will be anti-chirping. We can then perform follow up analysis on systems we determine to be mass transferring to search for the effects of novae. 

The analysis provided in this work is applicable beyond the case of mass transferring double white dwarf binaries. Any effects that induce sudden shifts in the frequency can be captured in the waveform studied, potential effects include tidal novae before mass transfer onset \cite{Fuller_2012}, compact object binaries with a white dwarf that enters tidal mode resonance \cite{Rathore:05}, or nova bursts in neutron star - white dwarf binaries \cite{Lewin:93}. The last case is much less likely than the DWD case due to fewer galactic binaries and a lower probability of prolonged stable mass transfer \cite{Bobrick:17}.

Performing the proposed analysis on future LISA data will help to resolve the effect of various size bursts on the orbit of close DWDs. As shown in section \ref{sec:Waveform}, an ejection of mass with conserved angular momentum leads to an expansion in the orbit. However, drag in the nova ejecta may overpower this effect and cause an overall reduction in the orbital separation \cite{Shen:15}. If this drag is the dominant effect, it may not be possible for long lived stably accreting DWDs to exist as novae drive the systems towards merger. Analysis of individual events as well as an estimate of the galactic event rate will provide a resolution to these uncertainties. The galactic DWD novae rate is of interest in its own right and can be determined with much higher accuracy through GW astronomy than is possible through electromagnetic observations as the majority of galactic novae occur in the galactic disk which cannot be well observed at large distances using current detection methods.

\section*{Acknowledgments}
This work is supported by Montana NASA EPSCoR Research Infrastructure Development under award No. 80NSSC22M0042 and by the National Science Foundation under award PHY-2308415 and CAREER award PHY-2541579. E.M. is supported by NASA FINESST award 80NSSC25K0310 and a Montana Space Grant Consortium Graduate Fellowship. NJC was supported by NASA LISA Preparatory Science Grant 80NSSC19K0320.
\appendix

\section{Posterior Volume} \label{adx:Posterior}

\begin{figure}
    \centering
    \includegraphics[width=0.8\linewidth]{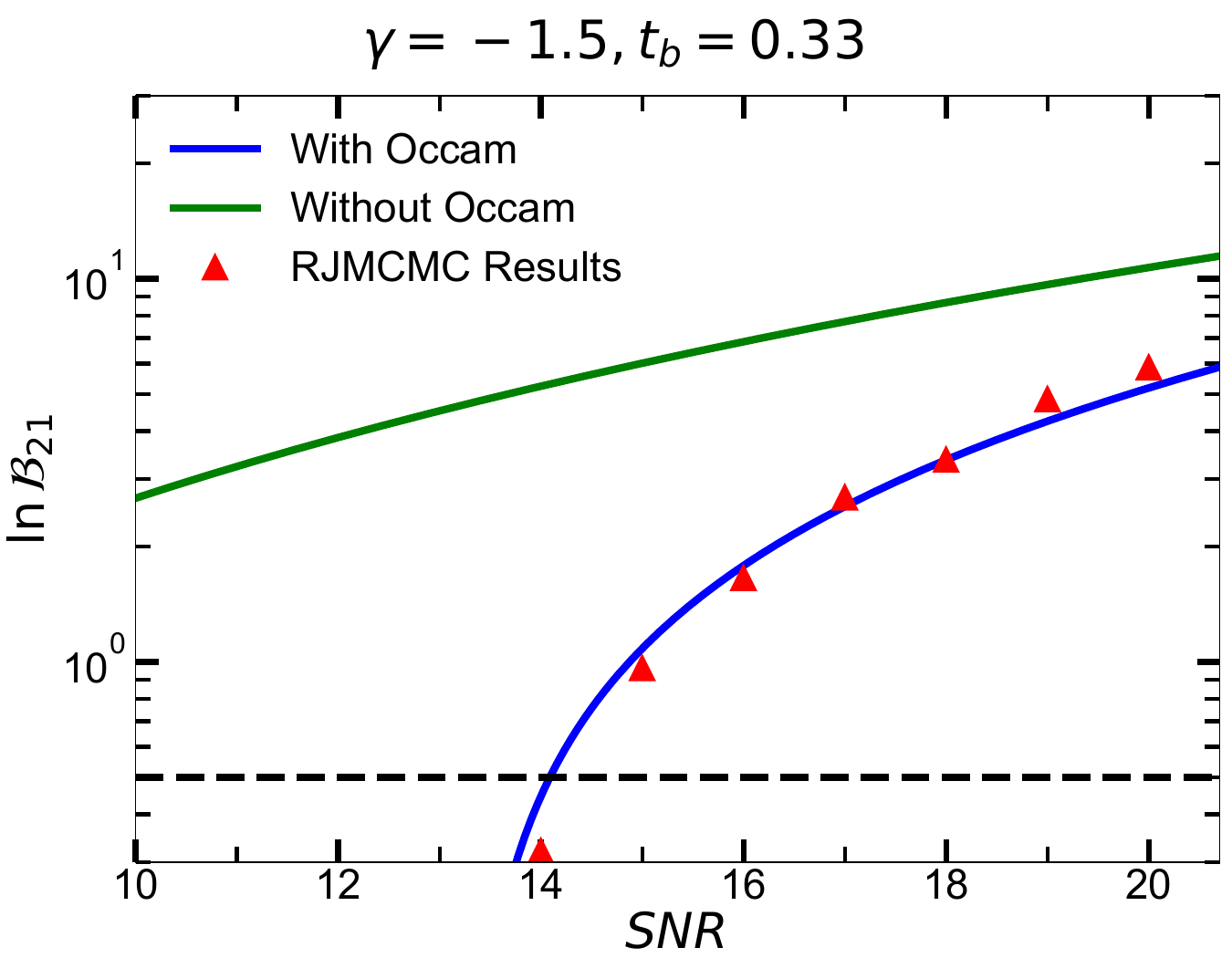}
    \caption{Results for Eq. \ref{eq:logBF expansion} with and without the Occam factor. We compare these results to Bayes factors obtained from our RJMCMC through computation of the odds ratio. The black dashed line shows $\ln \mathcal{B}_{21} = 0.5$. In this sequence of runs the priors on our burst parameters are limited to $|\gamma|/\alpha\in [10^{-7},10^{-3}] $ and $t_b \in [0,0.45]$ to prevent multi-modal posteriors.
    }
    \label{fig:multimodaltest}
\end{figure}

We show in Fig. \ref{fig:multimodaltest} by limiting the parameter space to eliminate multi-modality that the factor $C$ does in fact arise from the multi-modal shape of the posterior as an additional factor is no longer needed to reproduce results from a RJMCMC. For these results we limit the prior ranges on our burst parameters to $|\gamma|/\alpha\in [10^{-7},10^{-3}] $ and $t_b \in [0,0.45]$.

\section{Reversible Jump MCMC Overview} \label{adx: RJMCMC details}
We use a reversible jump MCMC in this work to transition between waveform templates with and without the effects of a nova-like explosion. Here we provide an overview of our jump probability following the helpful discussion in \cite{Sambridge:06}. 

We begin with Bayes's theorem \cite{Bayes:63}, which can be expressed as:
\begin{equation} \label{eqn:Bayes1}
    p(\bi{x}|\bi{d}) = \frac{p(\bi{d}|\bi{x})p(\bi{x})}{p(\bi{d})},
\end{equation}
where $p(\bi{x}|\bi{d})$ is the \emph{a posteriori} probability density of measuring unknowns $\bi{x}$ given some data $\bi{d}$, $p(\bi{d}|\bi{x})$ is the likelihood of observing the data $\bi{d}$ given specific values of $\bi{x}$, and $p(\bi{x})$ is the \emph{a priori} probability density of $\bi{x}$. The term $p(\bi{d})$ is the integrated probability of the likelihood weighted by the prior:
\begin{equation} \label{eqn:probint1}
    p(\bi{d}) = \int p(\bi{d}|\bi{x})p(\bi{x}) \rmd \bi{x}.
\end{equation}
For different expressions for $p(\bi{x})$ indexed by different numbers of unknowns $k$, we express equations \ref{eqn:Bayes1} and \ref{eqn:probint1} as 
\begin{equation} \label{eqn:Bayes2}
    p(\bi{x}|k,\bi{d}) = \frac{p(\bi{d}|\bi{x},k)p(\bi{x}|k)}{p(\bi{d}|k)}
\end{equation}
and
\begin{equation} \label{eqn:probint2}
    p(\bi{d}|k) = \int p(\bi{d}|\bi{x}, k)p(\bi{x}|k) \rmd \bi{x}.
\end{equation}
We can then use the property of probability distribution functions (PDFs) that
\begin{equation}
    p(x,y) = p(x|y)p(y)
\end{equation}
and use Bayes's theorem for $k$ to get
\begin{equation} \label{eqn:pxkd}
    p(\bi{x},k|\bi{d}) = \frac{p(\bi{d}|k,\bi{x})p(\bi{x}|k)p(k)}{p(\bi{d})}.
\end{equation}
This allows us to treat $k$ as an unknown in our analysis.

We are then able to use a Markov Chain Monte Carlo algorithm \cite{Smith:91, Gelfand:90, Smith:93} to determine the best values for $\bi{x}$ and $k$ given some data $\bi{d}$. To generate independent samples a random walk is performed by proposing moves from a current model $\bi{x}^{\rm p}$ to a proposed model $\bi{x}^{\rm q}$ which is accepted with probability $\eta$. For each step the proposed values $\bi{x}^{\rm q}$ are generated from the current values through some proposal distribution which is written as $q(\bi{x}^{\rm q}|\bi{x}^{\rm p})$. Then for a fixed-dimensional sampler the jump probability is given as
\begin{equation}
    \eta = {\rm Min} \left[1, \frac{p(\bi{x}^{\rm q}|k, \bi{d})q(\bi{x}^{\rm p}|\bi{x}^{\rm q})}{p(\bi{x}^{\rm p}|k, \bi{d})q(\bi{x}^{\rm q}|\bi{x}^{\rm p})}   \right].
\end{equation}
This is known as the Metropolis-Hastings rule \cite{Metropolis:49, Hastings:70}.

This rule can be applied to trans-dimensional jumps by generating random numbers $\bi{u}$ using a chosen distribution $g(\bi{u})$. We can then use a bijective mapping $h$ to calculate our proposed jumps
\begin{equation}
    \bi{x}^{\rm q} = h(\bi{x}^{\rm p}, \bi{u}).
\end{equation}
Green \cite{Green:95} showed that the Metropolis-Hastings jump probability becomes 
\begin{equation} \label{eqn:transalpha}
        \eta = {\rm Min} \left[1, \frac{p(\bi{x}^{\rm q}, k'| \bi{d})g'(\bi{u}^{\rm q})}{p(\bi{x}^{\rm p},k| \bi{d})g(\bi{u}^{\rm p})}|J|   \right],
\end{equation}
where $|J|$ is the Jacobian of the transformation:
\begin{equation}
   J = \left| \frac{\partial(\bi{x}^{\rm q},\bi{u}^{\rm q})}{\partial(\bi{x}^{\rm p},\bi{u}^{\rm p})} \right|.
\end{equation}
Now let's specify to the case where we are proposing a jump from a model without a burst to a model with one. In this case $k' = 6$, $k = 4$, so we need $k'-k=r=2$ random numbers. We choose our proposed parameter values as
\begin{eqnarray}
    x^{\rm q}_i = x^{\rm p}_i , i \in [1,4], \\
    x^{\rm q}_i = u^{\rm p}_{i-k} , i = 5,6.
\end{eqnarray}
This choice grants a simple Jacobian $|J| = 1$. We choose the distributions for $u^{\rm p}_{i}$ to match our burst parameter priors: $u^{\rm q}_{1}$ is chosen with a symmetric in sign uniform in log space distribution for fractional frequency shifts of $|\Delta f|/f =|\gamma|/\alpha\in [10^{-8},10^{-3}]$ and $u^{\rm q}_{2}$ is chosen as a uniform random number $U[0,1]$. We use equation \ref{eqn:pxkd} to rewrite the first fraction in equation \ref{eqn:transalpha} as
\begin{equation}
    \frac{p(\bi{x}^{\rm q}, k'| \bi{d})}{p(\bi{x}^{\rm p},k| \bi{d})} = \frac{p(\bi{d} | \bi{x}^{\rm q}, k') g(u_1^{\rm p}) P(k')}   {p(\bi{d} | \bi{x}^{\rm p}, k) P(k)},
\end{equation}
where $P(k)$ is the prior probability for each model.
This grants a jump probability 
\begin{equation} \label{eq:alphafinal}
    \eta = {\rm Min} \left[\frac{p(\bi{d} | \bi{x}^{\rm q}, k') P(k')}   {p(\bi{d} | \bi{x}^{\rm p}, k) P(k)}\right].
\end{equation}
For the reverse jump we choose our proposed parameter values as
\begin{equation}
    x^{\rm q}_i = x^{\rm p}_i , i \in [1,4].
\end{equation}
It is simple to show that the reverse jump grants the same probability, now with $k' = 4$ and $k = 6$. For most of this work we choose $P(\mathcal{H}_2)=P(\mathcal{H}_1)$ for simplicity. However, in generating Fig. \ref{fig:lnBFs} we choose $P(\mathcal{H}_1)/P(\mathcal{H}_2) = \exp[(\frac{1}{2}(1-FF^2)\rho^2)]$ to keep the odds ratio close to unity.
\vspace{1cm}

\bibliography{iopart-num.bib}
\end{document}